# Looplets: A Language For Structured Coiteration

Willow Ahrens
MIT CSAIL
Cambridge, MA, USA
willow.ahrens@mit.edu

Daniel Donenfeld
MIT CSAIL
Cambridge, MA, USA
danielbd@mit.edu

Fredrik Kjolstad
Stanford University
Stanford, CA, USA
kjolstad@cs.stanford.edu

Saman Amarasinghe
MIT CSAIL
Cambridge, MA, USA
saman@csail.mit.edu

## Abstract

Real world arrays often contain underlying structure, such as sparsity, runs of repeated values, or symmetry. Specializing for structure yields significant speedups. But automatically generating efficient code for structured data is challenging, especially when arrays with different structure interact. We show how to abstract over array structures so that the compiler can generate code to coiterate over any combination of them. Our technique enables new array formats (such as 1DVBL for irregular clustered sparsity), new iteration strategies (such as galloping intersections), and new operations over structured data (such as concatenation or convolution).

## 1 Introduction

Arrays (or tensors) are a powerful abstraction for representing collections of data. From scientific simulations to neural networks to image processing, array programming frameworks like NumPy, TensorFlow, and Halide help users productively process large amounts of data [2, 18, 42]. These complex frameworks are built on one of the simplest data structures in computer science: the dense array. A dense array stores every array element contiguously in memory. Iterating through a dense array is as easy as incrementing a pointer in a loop.

We can greatly improve on the efficiency of dense array processing by taking advantage of the underlying regular or irregular structure often present in real-world data. Data may be sparse (mostly zero), symmetric (mirrored along the diagonal), or contain repeated values. When data is sparse, we need only store the nonzeros. When data is symmetric, we need only store one half. When data contains runs of repeated values, we need only store one value from each run. With current memory sizes, many datasets are impossible to store without these optimizations. Storage improvements lead to performance improvements when we can use mathematical properties to avoid redundant work. For example, in a sparse sum, we need only add the nonzero values.

In these representations, iteration is more complicated than dense iteration. For example, to efficiently iterate over sparse arrays, we must skip over the zeros. Several sparse formats are used in practice for different situations, such as sorted lists, hash tables, or bytemaps, and special implementations must be developed for each. The iteration problem compounds when we need to combine sparse arrays that are stored in different formats. If we wish to multiply two sparse arrays pointwise, we must determine which nonzero locations are shared. The coiteration logic is a performance bottleneck, so we need custom implementations for the combinatorial space of potential input formats.

The influential TACO framework [26] compiles specialized code for all combinations of dense and sparse arrays. Users can also add custom sparse formats by implementing TACO's interface for iterators over nonzeros [11].

However, an iterator-over-nonzeros interface is not sufficient to express the full variety of underlying structures and iteration strategies encountered in real-world sparse arrays. Sparsity patterns may contain irregular clusters or blocks, or even regular shapes like banded or triangular matrices. When nonzeros are clustered, it is better to process the dense regions with simple dense code, rather than e.g. checking if each dense nonzero overlaps with nonzeros in other sparse arrays. TACO can model a dimension as entirely sparse (using the nonzero iterator interface) or entirely dense (using a random access interface), but cannot iterate over a single dimension as a combination of sparse and dense regions. TACO can artificially add dimensions to construct formats like BCSR (fixed-size blocks) or DIA (fixed number of diagonals), but these constructions cannot be composed unless both arrays share the same dimensional transform.

We must also move beyond mere irregular sparsity, to also express structure. Code that iterates over nonzeros does not sufficiently accelerate computations over symmetric arrays or arrays with repeated values, since these may not contain any zeros at all. Therefore, a myriad of new compilers or compiler passes have been developed in separate attempts to support different structural specializations, such as ragged arrays [17], symmetry [45], or run-length encoding [14]. These extensions represent significant implementation effort, but do not compose with one another.

In this paper, we show how to abstract over these iteration strategies using the concept of iterator protocols. An iterator **protocol** describes the interface that a structured array format must implement to be used in some particular implementation of an array operation. Protocols are abstract representations of structure within a sequence of values. The protocol declares which functions a format should implement, and how to interpret the data those functions produce. For example, the iterator-over-nonzeros protocol asks formats to implement iterators over coordinates and corresponding values, and declares that we should interpret the format as a vector which is zero except at the coordinates.

In short, dense array programming is comparatively easy to implement because it only needs to support one protocol: unstructured random access. The TACO compiler supports two protocols, random access and an iterator-over-nonzeros. Ragged arrays require a protocol for dense regions followed by zeros [17], and TACO's run-length encoding extension supported a protocol for sequences of runs. The systems do not compose because each system hard-coded support for its respective protocol. As applications for array compilers become more diverse, we must support an ever-increasing number of protocols and the interactions between them. Hand-writing code for each situation is infeasible.

We introduce a language for iterator protocols. We express iterators over arrays using basic functional units called **Looplets** that expose underlying array structure in a customizable way. We then devise a compiler that can automatically generate efficient code for combinations of different input protocols and array expressions. We integrate the Looplet language into a new array compiler called **Finch**, which accelerates the design space exploration of sparse and structured loop nests. Finch can compile expressions over any combination of structured formats, protocols, mathematical simplifications, and index modifiers. Finch lowers expressions progressively, allowing us to express structural simplifications (like zero-annihilation from sparsity) through straightforward and extensible rewrite rules.

A composable protocol language makes it possible to express far more data representations than prior work [11] and, critically, to mix them freely in expressions. Protocols can be easily developed for new formats, such as the PackBITS format, which intersperses runs of repeated values with dense regions of unstructured data and is standardized as part of the TIFF image format [1]. Different protocols can also be developed for the same format. For example, a list of nonzeros might be better traversed by skipping ahead with a binary search. If there are two lists, we might want one to lead and the other to follow, or perhaps allow each to skip ahead to nonzeros in the other, a "galloping" intersection strategy [5]. Galloping is used to implement worst-case-optimal joins in databases [40, 52]. Finally, protocols can be modified to implement more complex operations. We can express padding operations by nesting an existing protocol within another, or affine indexing by shifting an existing protocol. These protocol modifications enable operations like concatenation or convolution over structured formats.

We make the following contributions:

- We propose the use of access protocols to abstract over array iteration strategies and introduce the Looplet language to express access protocols.
- We describe a compiler for the Looplet language that combines multiple local protocols into one efficient loop nest that implements an array operation.
- We show how a surprising range of iteration strategies, formats and array operations can be expressed and composed using Looplets.

To evaluate our contributions, we benchmark Finch on a diverse set of operations; sparse matrix sparse vector multiply, triangle counting, sparse input convolution, alpha blending, and all-pairs image similarity. We compare to OpenCV and TACO, and our evaluation shows that Finch is competitive on the kernels that each supports, while allowing them to be composed. Different protocol perform better in different cases, emphasizing the need for flexibility in iteration strategies. In some cases, Finch obtains order-of-magnitude speedups over the state of the art.

## 2 Motivating Example

When there is only one input to an operation, such as map or reduce, it is comparatively easy to hand-write specialized implementations for each data structure that the input may be stored in. However, when there are multiple inputs and multiple operations, we cannot easily hand-write all combinations of data structures and operations. Even when both inputs are sparse, there is no universally efficient protocol. Here, we demonstrate how the iterator-over-nonzeros interface of systems like TACO [26] cannot express the appropriate coiteration over unstructured and clustered sparsity.

Consider the dot product kernel, which combines two vectors by summing their pairwise product. The dot product can be written as $C = \sum_i A_i B_i$. When both vectors are dense, we compute it with a single for-loop. When both vectors are sparse, we might consider using TACO's two-finger merge template, which represents both vectors as iterators-over-nonzeros, and merges the nonzero coordinates to find shared nonzeros. However, this is not a good fit when one or both of the vectors are clustered.

Perhaps the quintessential sparse format is the **sparse list** format (referred to as "compressed" by Kjolstad et. al. [26]), which is a good fit for unpredictable sparsity patterns. This format stores the locations of only nonzero values using a sorted list, which is a natural fit for an iterator-over-nonzeros interface. However, when nonzeros are clustered they benefit from more specialized formats. For instance, banded matrices are popular in scientific computing and contain dense regions of nonzeros centered around the diagonal. To represent such structures, we can introduce a **sparse band** format that stores a single, variably wide block of contiguous nonzeros. The left columns of Figures 1a–1c express coiteration between these sparsity structures in the compressed iterator-over-nonzeros model, and the result of inlining our interface definitions into a two-finger-merge dot-product template. The resulting code iterates over the nonzero values of both lists, stopping once either is exhausted.

The right columns of Figures 1a–1c express coiteration in the Looplets model. Pipeline, phase, stepper, spike, run and lookup are all Looplet types that will be explained in Section 3, but they describe the semi-structured sparsity of the data in Figure 1c. The increased expressiveness of the model lets us inform the compiler that there are large zero regions before and after the dense band. This lets us skip ahead in the sparse list to the start of the dense band region. Additionally, our protocol declares that the dense band can be randomly accessed. This allows us to skip the elements in the band that are zero in the list. These two optimizations can have asymptotic effects, which are visualized for our example vectors in Figure 1c.

```
getnnz(A::SpList) = length(A.idx)
getidx(A::SpList, p) = A.idx[p]
getval(A::SpList, p) = A.val[p]
```

```
unfurl(A::SpList) =
Pipeline(Phase(Stepper(Spike(...))),
↪  Phase(Run(0)))
```

```
getnnz(A::SpBand) = A.stop-A.start
getidx(A::SpBand, p) = A.start+p-1
getval(A::SpBand, p) = A.val[p]
```

```
unfurl(A::SpBand) =
Pipeline(Phase(Run(0)),
↪  Phase(Lookup(...)), Phase(Run(0)))
```

**(a)** Comparing iteration interfaces. On left, an iterator-over-nonzeros implementation of a sorted coordinate list and our banded format. On right, equivalent Looplet declarations (simplified from Figures 3d and 3f) that expose more structure to the compiler.

```
function dot(A::SpList, B::SpBand)
  C = 0
  pA = 1
  PA = length(A.idx) #getnnz(A)
  pB = 1
  PB = B.stop-B.start #getnnz(B)
  while pA <= PA && pB <= PB
    iA = A.idx[pA] #getidx(A, pA)
    iB = B.start+pB-1 #getidx(B, pB)
    i = min(iA, iB)
    if i == iA && i == iB
      vA = A.val[pA] #getval(A, pA)
      vB = B.val[pB] #getval(B, pB)
      C += vA * vB
    end
    pA += iA == i
    pB += iB == i
  end
  return C
end
```

```
function dot(A::SpList, B::SpBand)
  C = 0
  i = B.start
  phase_stop = min(B.stop, A.idx[end])
  if phase_stop >= i
    pA = search(A.idx, i)
    iA = A.idx[pA]
    while i <= phase_stop
      if iA <= phase_stop
        i = iA
        vA = A.val[pA]
        vB = B.val[(i - B.start) + 1]
        C += vA * vB
        pA += 1
        iA = A.idx[pA]
      else
        i = phase_stop
      end
      i += 1
    end
  end
end
```

**(b)** The resulting dot-product code from iterator-over-nonzeros (left) and Looplets (right). On left, a (straightforward) Julia translation of TACO output, where we have replaced the TACO compressed level functions with that of our hypothetical banded matrix format. On right, simplified Finch output.

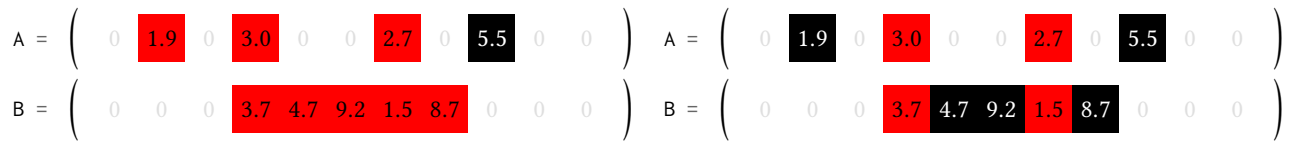

**(c)** An example execution of each algorithm. The nonzero locations processed by each dot product inner loop are shown in red, unprocessed nonzeros are shown in black. The iterator-over-nonzeros code (left) processes nonzeros from both lists till one is exhausted. The looplet code (right) skips to the start of the block, then randomly accesses it, thus improving asymptotic efficiency.

**Figure 1.** Coiteration comparison between an iterator-over-nonzeros approach (left) and our Looplets approach (right) to coiteration over a sparse list and sparse band format. The list format holds many scattered nonzeros, while the band format holds a single dense nonzero region. Elsewhere we describe our VBL format that holds multiple bands.

## 3 Looplet Language

Looplets are abstract descriptions of regular or irregular patterns in a sequence of values, together with the code needed to iterate over the sequence. Looplets represent sequences using hierarchical descriptions of the values within each region or subregion. Regions are specified by their absolute starting and ending index, referred to together as an **extent**.

Looplets are lowered by a compiler. Values can be static (i.e., known at compile time), or dynamic (i.e., known only at runtime). Looplets represent the full sequence abstractly, even if all of the values are not stored explicitly. For example, a **run** looplet represents a sequence of many of the same value, usually stored once. A **lookup** looplet represents an arbitrary sequence as a function of the index. While this function is often an array access, it could also be a function call, like $f(i) = sin(\pi i/7)$.

Looplets are defined with respect to the extent of the **target region** that we wish to represent (usually the range of an enclosing loop). The **spike** looplet represents a sequence of values that are all the same, followed by a single scalar value at the end of the target region. Frequently, we will want to represent a subregion, or **truncation**, of a looplet. Many looplets are self similar. A run looplet, for example, can represent any subregion of itself. Other looplets are specific to the target region. For example, a truncation of a spike that excludes the last element produces a run.

Some looplets represent the composition of other looplets. The **switch** looplet represents a choice between different looplets under different conditions. The **pipeline** looplet represents a sequence of a few different looplets, one after the other. The **stepper** looplet represents a sequence of an unbounded number of identical looplets. These looplets describe not only their sublooplets, but also the conditions and extents in which sublooplet apply.

A looplet nest can introduce and manipulate its own runtime variables in the generated code. For example, the code to advance a stepper to the next looplet in a sequence might increment some counter variable. A switch looplet might **`Lookup(body(idx))`.** An arbitrary sequence of scalars where the element at index i can be computed as `body(i)`.

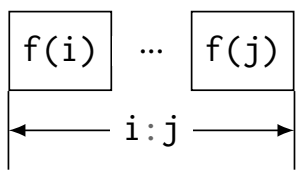

**`Run(body)`.** A sequence of the same repeated scalar body.

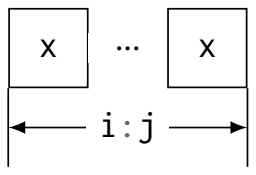

**`Spike(body, tail)`.** A sequence of the same repeated scalar body, followed by some scalar `tail` at `stop`.

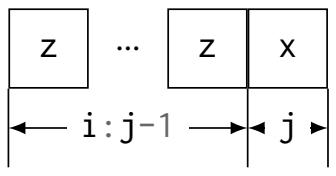

**`Pipeline(Phase(stride, body(ext)), ...)`.** The concatenation of a few different child looplets in a sequence. Each phase declares its corresponding extent, ending at `stride`, and the child `body(ext)` for some target subextent.

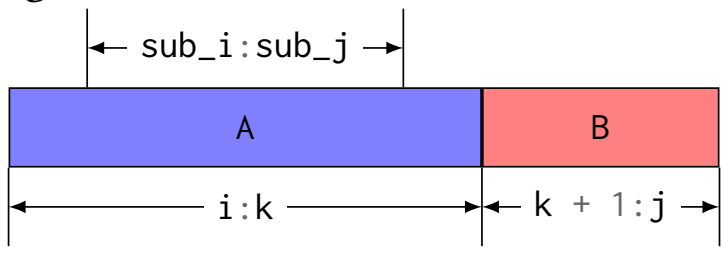

**`Stepper(seek(idx), stride, body(ext), next)`.** The repeated application of the same child looplet `body(ext)`, each child ending at `stride`. Steppers are evaluated iteratively, so `next` advances state to the next looplet and `seek(i)` initializes variables for a starting index `i`.

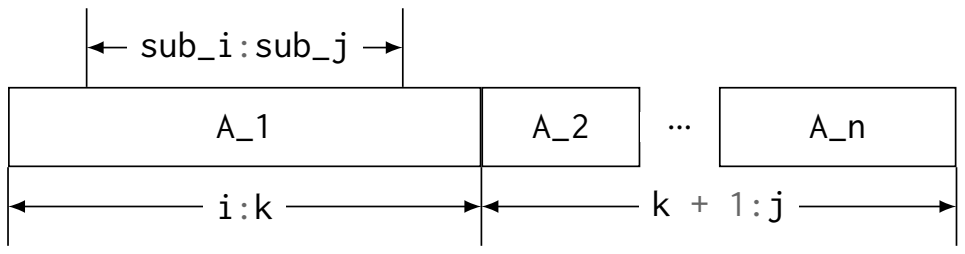

**`Jumper(seek(idx), stride, body(ext), next)`.** Like a stepper, but `ext` may be wider than the extent declared by `stride`, enabling accelerated iteration (e.g. galloping).

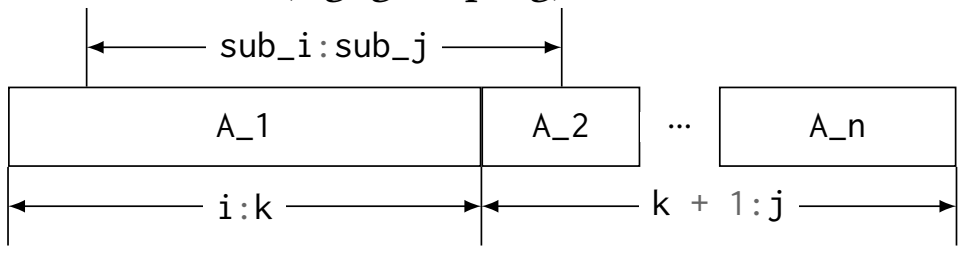

**`Shift(delta, body)`.** A wrapper that shifts all declared extents of body by `delta`. Shifting is necessary because extents are absolute, rather than relative.

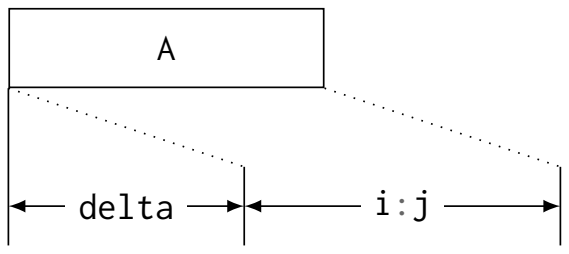

**`Switch(Case(cond, body), ...)`.** Represents the first child looplet body for which `cond` evaluates to true at runtime.

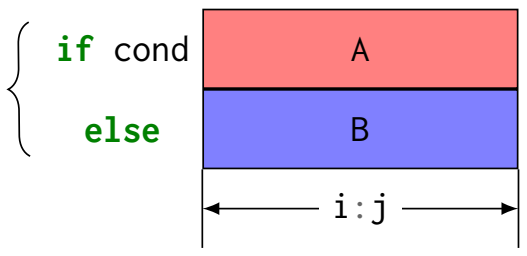

**Figure 2.** The looplets considered in this paper, described and displayed with a target extent of `i:j`.

use that variable in its condition. Looplets are executed in ascending index order, but some regions may be skipped.

Precise descriptions of all the looplets are given in Figure 2.

## 4 Formats

Array structures are diverse, and so are approaches for array storage. Prior approaches, such as TACO level formats [11] and the FiberTree abstraction [49], popularized hierarchical taxonomies for array structure. These abstractions decompose multidimensional arrays mode-by-mode into trees, where each mode is a level of the tree, and each node within a level corresponds to a slice. For example, the popular CSR matrix format stores a list of nonzero columns within every row. This can be expressed with an dense outer row level and a sparse list inner column level.

It is helpful to view an array $A$ as a function mapping several indices to elements. If we were to curry our array function, then a partial application corresponds to a slice of the array. For example, if $A$ were a 3-dimensional, then $A(i_1, i_2, i_3) = A(i_1)(i_2)(i_3)$ and $A(i_1) = A[i_1, :, :]$. In this paper, we define a **fiber** as an array function mapping a single index to a subfiber. Fibers can be thought of as abstract vectors of subfibers. The **level storage** is a datastructure responsible for storing all the fibers within a dimension.

Looplets make it easy to efficiently support new level formats. The array tree abstraction decomposes multidimensional array formats into unidimensional ones. Looplets further decompose the remaining unidimensional structure. The format developer can use looplets to describe the structure of a single fiber within a level. In Section 6, we show how the looplet nest is then automatically merged with other nests and lowered to efficient code. We call the process of constructing a looplet nest for a fiber as **unfurling**.

Since each level may store many fibers, we also define an **environment storage** as the internal state used to identify which fiber is represented. The same Looplet nest is often used to process different fibers within the same level. The environment storage holds static or dynamic information about the path taken from the root of the array tree to get to the fiber in question. We can also use the environment to pass other information to child levels, such as cached results or requests to compute statistical information. We can also express a level that spans multiple modes by modifying the environment without returning a new sublevel.

By defining just a few level formats, we can express a wide variety of structured array storage formats. Figure 3 gives some examples of level formats for a diversity of matrix structures, along with one or more protocols that allow us to efficiently traverse them.

Our approach relaxes the constraints of prior work on array structure. Rather than force array datastructures to express a set of points that are nonzero, our format can express any structured function from index to value. Our relaxed approach to hierarchical storage is enabled by the looplet abstraction, but not required. Other array storage approaches might also be implemented with looplets. For example, an external standard library format could express protocols using looplets to compose with our framework.

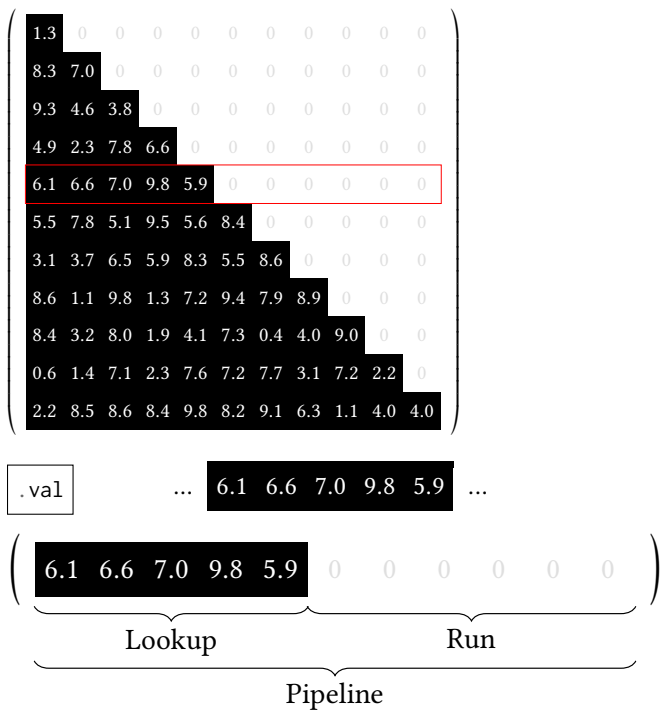

```
offset = i*(i-1)/2
Pipeline(
  Phase(
    stride = i,
    body = Lookup(
      body(j) = val[offset+j])),
  Phase(
    body = Run(0)))
```

**(a)** Lower Triangular

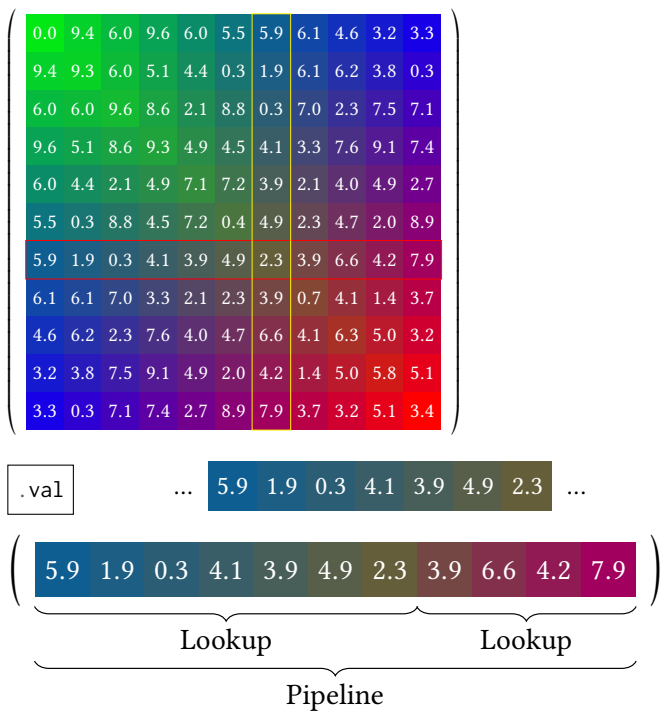

```
offset = i*(i-1)/2
Pipeline(
  Phase(
    stride = i,
    body = Lookup(
      body(j) = val[offset+j])),
  Phase(
    body = Lookup(
      body(j) = val[j*(j-1)/2+i])))
```

**(c)** Symmetric

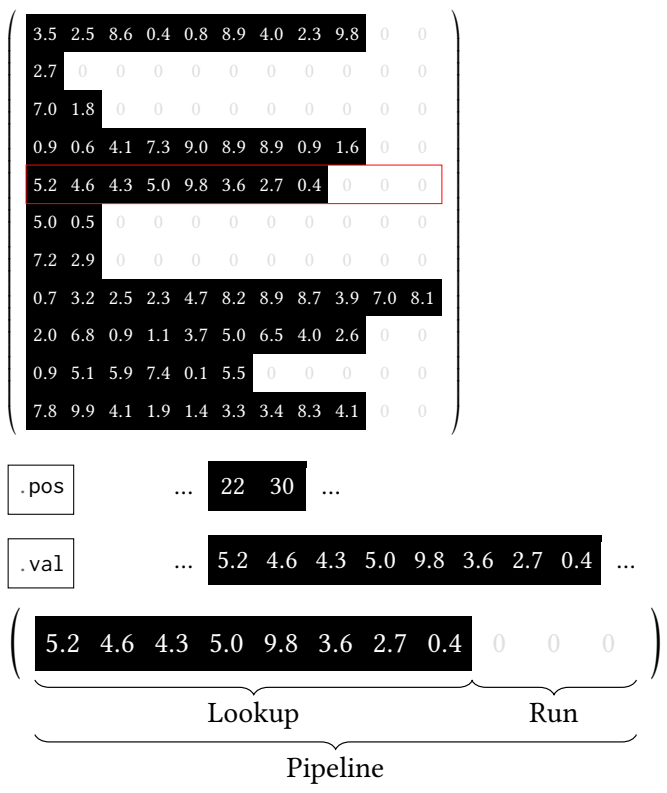

```
Pipeline(
  Phase(
    stride = pos[i+1]-pos[i],
    body = Lookup(
      body(j) = val[pos[i]+j-1])),
  Phase(
    body = Run(0)))
```

**(e)** Ragged

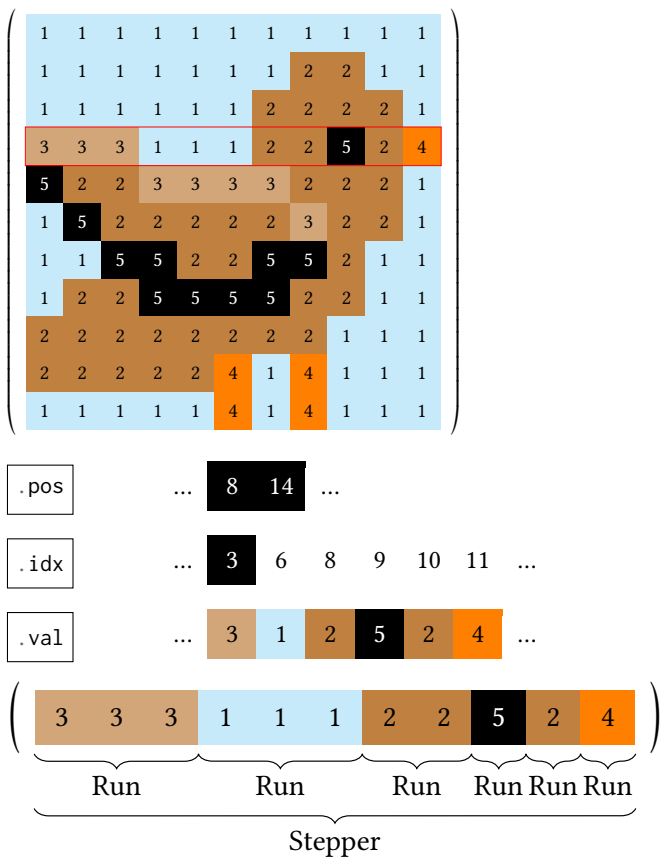

```
p = pos[i]
Stepper(
  seek(j) = (p = search(idx, j)),
  stride = idx[p],
  body = Run(
    body = val[p]),
  next = p += 1)
```

**(g)** Image with repeated values. Run-Length Format.

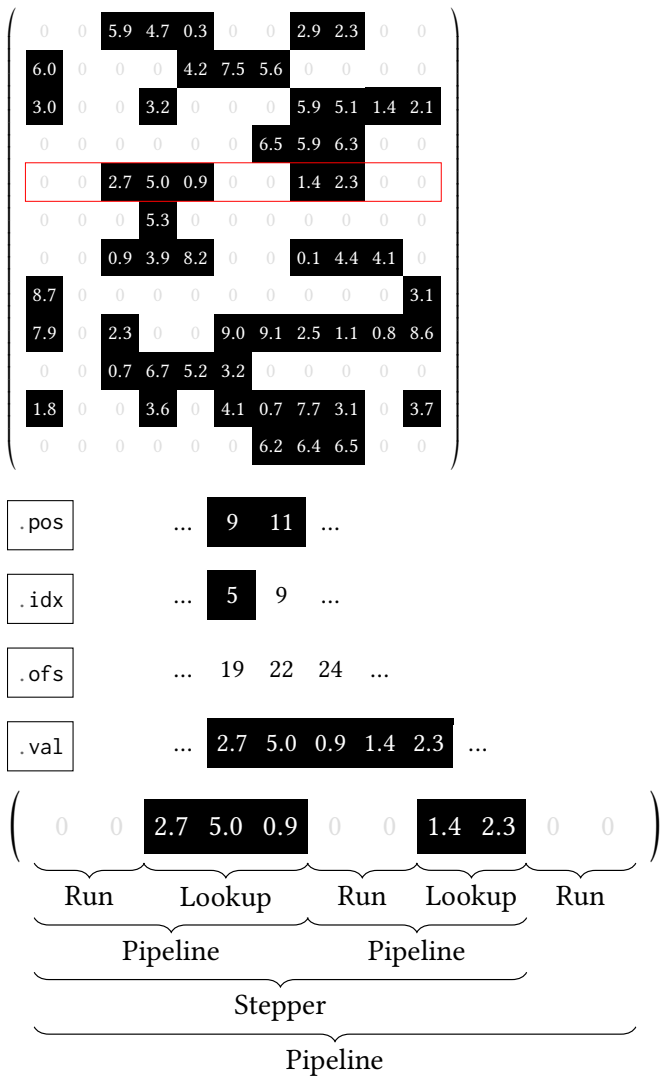

```
Pipeline(
  Phase(
    stride = idx[pos[i+1]-1],
    body = begin
      p = pos[i]
      Stepper(
        seek(j) = (p = search(idx, j)),
        stride = idx[p],
        body = Pipeline(
          Phase(
            stride =
↪   idx[p]-(ofs[p+1]-ofs[p]),
            body = Run(0)),
          Phase(
            body = Lookup(
              body(j) =
                val[ofs[p+1]+j-idx[p]]))),
        next = p += 1))
  Phase(
    body = Run(0)))
```

**(b)** Clustered Matrix, VBL Format

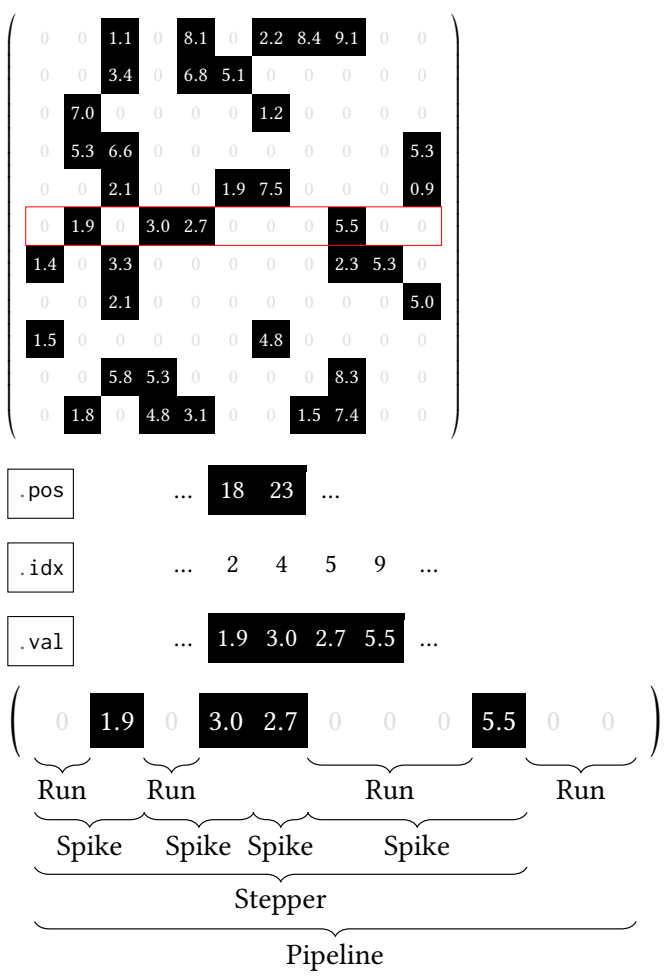

```
Pipeline(
  Phase(
    stride = idx[pos[i+1]-1],
    body = begin
      p = pos[i]
      Stepper(
        seek(j) = (p = search(idx, j)),
        stride = idx[p],
        body = Spike(
          body = Run(0)
          tail = val[p]),
        next = p += 1)
    end)
  Phase(
    body = Run(0)))
```

**(d)** Uniform Sparsity, List Format

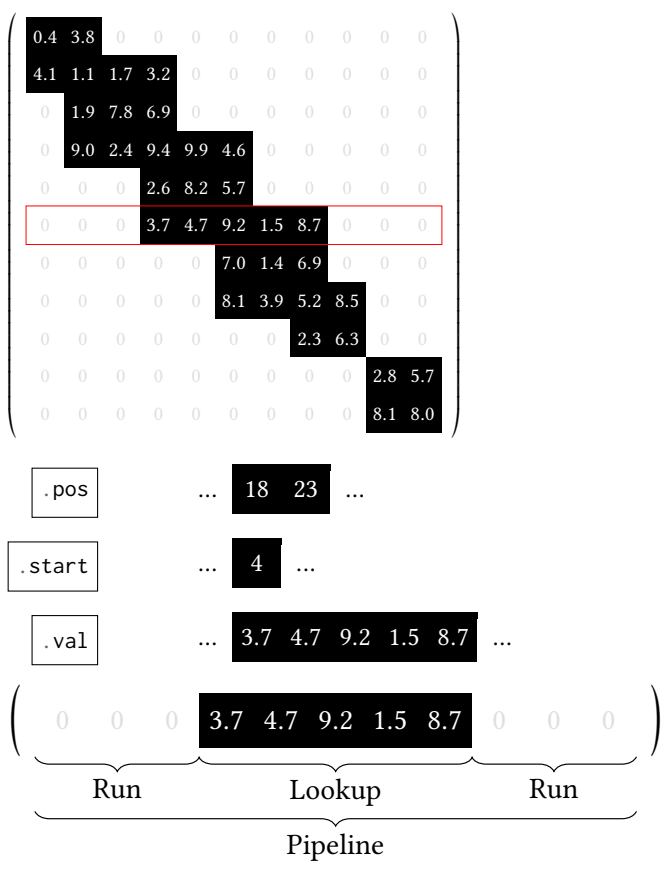

```
band = pos[i + 1] - pos[i]
offset = pos[i] - start[i]
Pipeline(
  Phase(
    stride = start[i] - 1
    body = Run(0)),
  Phase(
    stride = start[i] + band - 1,
    body = Lookup(
      body(j) = vals[offset + j])),
  Phase(
    body = Run(0)))
```

**(f)** Banded Matrix, Band Format

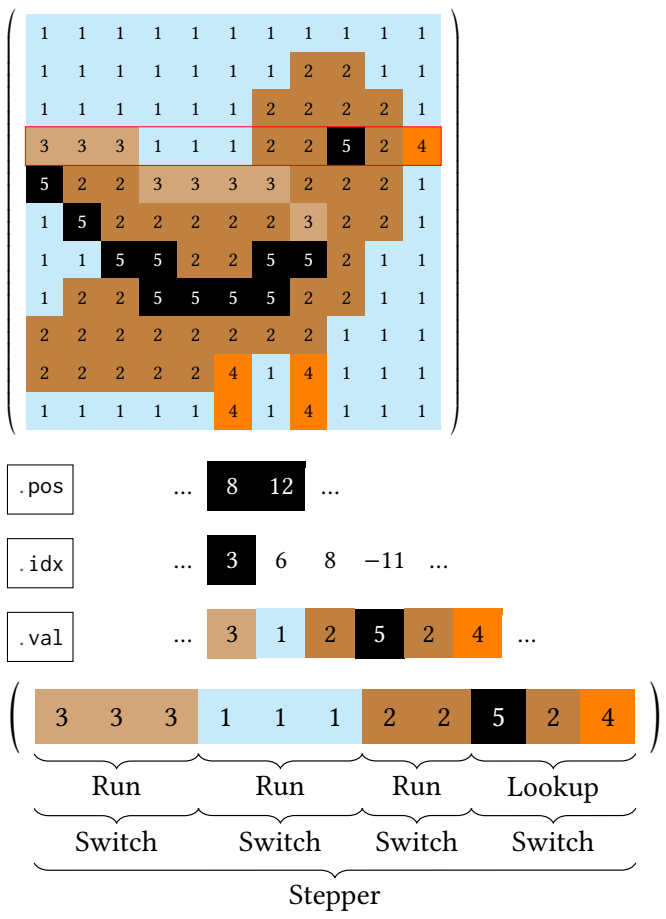

```
s = 0
p = pos[i]
Stepper(
  seek(j) = (
    p = search(abs(idx), j)),
  stride = idx[p],
  body = Switch(
    Case(
      cond = idx[p] > 0,
      body = Run(
        body = vals[p]))
    Case(
      body = Lookup(
        body(j) = vals[p + j - s])))
  next = s = abs(idx[p]); p += 1;)
```

**(h)** Image with repeated values. PackBITS Format.

**Figure 3.** A variety of example structures, corresponding level formats, and protocols, followed by their corresponding looplet nest unfurling code. Matrices are row major, and outer levels are dense. The row under consideration is highlighted in red.

## 5 Extended Concrete Index Notation

Concrete index notation (CIN) is a high level language for array computations. CIN was first introduced as part of the TACO compiler [25]. It was later extended to include **multi** statements (to express multiple outputs) [27] and protocol declarations (to customize the access protocol independently from the array format) [3].

### 5.1 Concrete Index Notation Background

Our new grammar for CIN is shown in Figure 4. At the heart of CIN is the assignment statement, which represents an update to a single array element. We can either increment or overwrite the element by a pointwise expression composed of functions over other array access expressions. Arrays are accessed by index variables. The forall statement repeats the assignment statements for each value of an index variable. The where statement lets us compute an array in one statement, then use it in the other. The producer (right hand) side computes arrays for the consumer (left hand) side to use. This gives rise to the notion of array results. Each statement in concrete index notation returns one or more results. The result of an assignment is the array being modified, and the result of a forall is the result of its body. The result of a where statement is the result of its consumer. Result arrays are initialized as soon as they enter the scope, and are finalized when they leave scope. Thus, arrays are initialized in the outermost where statement which contains them on the right hand side, or at the start of the program. The multi statement allows us to compute multiple outputs at once, and it combines the results of its constituent statements.

### 5.2 Concrete Index Notation Extensions

We extend CIN with additional constructs to express looplets. We allow arrays to be accessed by any expression, not just indices. Critically, index expressions allow us to specify different **protocols** for accessing index variables. Instead of returning a fixed looplet nest, users can specify the kind of nest that should be used. For example, a user might choose between random-access, iterating over nonzeros, or galloping over some index variable. Index expressions also enable **index modifiers**, which affect the protocol of the index they modify. Users might declare that they should iterate over a slice of an array, or perhaps shift the index by some expression. If an index expression is opaque to the compiler, it represents a scatter operation. To support scatters, we introduce the **sieve** statement. A sieve only iterates over the subset of loop iterations where its condition is true. Thus, we can transform random accesses into a sequential accesses, using a special mask protocol (`Pipeline(Run(false), true, Run(false))`) to efficiently represent `j == f(i)`.

```
@∀ i A[i] = B[f(i)] \to @∀ i j @sieve j == f(j) A[i] = B[j]
```

During lowering, our code generator may need to introduce **pass** statements, which are no-ops that do nothing other than remember which array outputs they aren't writing to. Pass statements return their arrays unmodified.

```
    expr = literal     call = expr + expr       assign = ACCESS = expr
           index              expr * expr                ACCESS += expr
           call               expr(expr...)              ACCESS *= expr
           access                                        ACCESS <<value>>= expr
           proto     access = VALUE[expr...]
           (expr)                                extent = expr : expr
           $value    proto  = expr::value        forall = @∀ index stmt
                                                          @∀ index ∈ extent stmt
 literal = 42        stmt   = assign
           3.14               forall             _where = (STMT) where (stmt)
           ...                _where
                              multi               multi = @multi STMT...
   index = i                  sieve
           j                  pass                sieve = @sieve expr STMT
           ...                (stmt)
                              $value               pass = @pass value...
```

**Figure 4.** Our extended concrete index notation grammar. Here, `value` is used for values (like arrays) in the scope surrounding the program, and `$(value)` is an escape sequence. The token `@finch` is used to denote a CIN program execution within a larger piece of code.

## 6 Lowering

Our compiler, Finch, lowers concrete index notation recursively, node by node, emitting code until a forall is reached. At a forall, Finch unfurls all arrays accessed at the outermost level by the forall index. This section describes our looplet merging and lowering algorithm.

We take a progressive lowering approach to looplet evaluation. Each type of looplet has a corresponding compiler pass that is capable of evaluating it. Many operations produce subexpressions that are lowered recursively. We lower whichever looplets we can at each step. Unlowered looplets are truncated or ignored for later stages, as applicable.

### 6.1 Looplet Lowerers

Since looplets are defined with respect to a forall statement, its index, and its bounds, each looplet lowerer operates on a forall loop expression. The looplets themselves are treated as vectors being accessed in the body of the forall loop.

***Lookups.*** The simplest way to evaluate a forall loop is to execute each iteration in turn, and evaluate the body of the loop after binding the index value. If all of the looplets in the body are lookups, dynamic values, or scalars with respect to the loop index, Finch emits a for-loop and lower the body after substituting a dynamic index value:

```
A = Lookup(body(j) = j^2)
B = Lookup(body(j) = data[j])
@finch(@∀ i ∈ 1:I (C[] += 2 * x * A[i] * B[i]))
⇓
for i_1 = 1:I
  @finch(C[] += 2 * x * $(i_1^2) * $(data[i_1]))
end
```

***Runs and Rewriting.*** When the loop body contains runs, Finch unwraps the runs into their scalar values and simplifies the resulting program. For example, sparse computing relies

```
f(a...) =>  if (f, a...) isa Constant eval(f(a...)) end

@loop $i @pass(a...) => pass(a...),
$a where @pass() => a,

+(a..., +(b...), c...) => +(a..., b..., c...),
+(a..., 0, b...) => +(a..., b...),
a[i...] += 0 => @pass(a),
a - b => a + (- b),
- (- a) => a,
*(a..., *(b...), c...) => *(a..., b..., c...),
*(a..., 1, b...) => @i *(a..., b...),
*(a..., 0, b...) => 0,
(*)(a) => a,
(*)(a..., - b, c...) => -(*(a..., $b, c...)),
a[i...] *= 1 => pass(a),
@sieve true $a => a,
@sieve false $a => pass(getresults(a)...),

or(a..., false, b...) => @i or(a..., b...),
or(a..., true, b...) => true,
or() => false,

f(a..., missing, b...) => missing,
a[i..., missing, j...] => missing,
coalesce(a..., missing, b...) => coalesce(a..., b...),

@loop i ∈ start:stop A[j] <<min>>= b => if j != i A[j] <<min>> = b end
@loop i ∈ start:stop A[j] += b => if j != i A[j] += b*(stop-start) end
```

**Figure 5.** A selection of rewrite rules used in Finch to declare mathematical properties enabling sparse and structural optimizations. Users can add custom rules for the kinds of computations in their domain.

on the fact that $x \cdot 0 = 0$. When one array is zero within a region, implementations can ignore arrays it multiplies.

Because Finch lowers separate looplet expressions for each subregion, simplifying optimizations can be expressed as rewrite rules. Using rewrite rules to express optimizations broadens the accessibility of our system. Users might express arbitrary rewrites for the interaction between custom value types and custom functions, such as semirings or beyond [13, 23, 37]. Figure 5 gives some examples of the kinds of rules we use. In addition to simple rules like zero-annihilation or constant propagation, some rules might operate on statements within the expression. For example, Finch recognizes that adding a constant $c$ to the same output $n$ times is equivalent to adding $c \cdot n$, saving $O(n)$ work. Removing loops over constants is useful for optimizing operations over run-length-encoded data. Users can even write their own simplifying compiler passes over intermediate expressions. Simplification passes should happen as early as possible in the lowering hierarchy. The Finch implementation recognizes a no-op `Simplify` looplet, which triggers a simplification pass.

```
A = Run(body = x)
B = Run(body = 0)
@finch(@∀ i ∈ start:stop (C[] += A[i] * B[i]))
⇓
@finch(@∀ i ∈ start:stop (C[] += 0 * $x))
⇓
@finch(@∀ i ∈ start:stop @pass C)
⇓
@finch(@pass C)
```

***Spikes.*** Given an expression with spikes, Finch constructs two subexpressions, one for the spike bodies (runs of variable length) and one for the tails (of length one). Finch truncates other looplets to match each case. When a loop has length one, Finch skips the loop and just evaluates the body once.

Recall that spike looplets depend on the target region. Thus, when spikes are truncated by other looplet passes, they produce a cases statement depending on whether the new target range includes the final tail element. If that last element is included, the subrange still represents a spike. Otherwise, it is simplified to a run:

```
A = Spike(body = 0, tail(j) = Adata[j])
B = Spike(body = 0, tail(j) = Bdata[j])
@finch(@∀ i ∈ start:stop (C[] += A[i] * B[i]))
⇓
@finch(@∀ i ∈ start:(stop - 1) (C[] += 0 * 0))   #body region
@finch((C[] += $(Adata[stop]) * $(Bdata[stop]))) #tail region
⇓
@finch((C[] += $(Adata[stop]) * $(Bdata[stop])))
```

***Switches.*** The switch lowerer produces a separate expression for each combination of cases from the switch looplets in an expression. Each combination is lowered separately and emitted in an if-else-block:

```
A = Switch(
  Case(cond = :(x > 1), body = 1),
  Case(cond = :(true), body = 2),
)
B = Switch(
  Case(cond = :(y > 1), body = 3),
  Case(cond = :(true), body = 4),
)
@finch(@∀ i ∈ 1:I (C[] += A[i] * B[i]))
⇓
if x > 1 && y > 1
  @finch(@∀ i ∈ 1:I (C[] += 1 * 3))
elseif x > 1
  @finch(@∀ i ∈ 1:I (C[] += 1 * 4))
elseif y > 1
  @finch(@∀ i ∈ 1:I (C[] += 2 * 3))
else
  @finch(@∀ i ∈ 1:I (C[] += 2 * 4))
end
```

***Pipelines.*** The pipeline lowerer produces a separate expression for each combination of phases from the pipeline looplets in an expression. The ranges of all the phases in each combination are intersected, and other looplets are truncated to match. Note that many of these combinations will have an empty intersection. Consider the graph where phase combinations are nodes and edges represent transitions from one phase to the next within a pipeline. If we lower combinations of phases in an order that linearizes the graph, then earlier phase combinations will always be executed before later ones. Since each edge advances a pipeline, the graph is acyclic and we can construct such an order:

```
A = Pipeline(
  Phase(stride = s_A, body = 1),
  Phase(body = 2),
)
B = Pipeline(
  Phase(stride = s_B, body = 3),
  Phase(body = 4),
)
@finch(@∀ i ∈ start:stop (C[] += A[i] * B[i]))
⇓
@finch(@∀ i ∈ start:min(s_A, s_B, stop) (C[] += 1 * 3))
@finch(@∀ i ∈ max(start, s_B):min(s_A, stop) (C[] += 1 * 4))
@finch(@∀ i ∈ max(start, s_A):min(s_B, stop) (C[] += 2 * 3))
@finch(@∀ i ∈ max(start, s_A, s_B):stop (C[] += 2 * 4))
```

***Steppers.*** Steppers represent an arbitrary number of child looplets. Finch first uses the stepper `seek` function to set each stepper's current child to intersect with the starting index of the current target region. Finch then uses a while loop to lower steppers, with each step evaluating as large a range as possible without crossing any stepper's child boundaries. At the beginning of the loop body, Finch computes the target region by intersecting the extent of each steppers' current child. Then, Finch truncates the stepper children to the computed region, producing the loop body expression of the step. Finally, Finch emits the `next` statements from each stepper, which are responsible for advancing the state of the child looplet to represent the next child, if necessary.

Because the while loop maintains the current starting index for the step, only the ending index is needed from each stepper. This single index is the `stride`. The `seek` function often contains a binary search, and the `next` function usually increments a variable, but they can be more general if needed:

```
A = Stepper(stride = idx[p], body = Run(val[p]), next = p += 1)
B = Stepper(stride = jdx[q], body = Run(wal[q]), next = q += 1)
p = q = 1
@finch(@∀ i ∈ start:stop (C[] += A[i] * B[i]))
⇓
p = q = 1
step = start
while step < stop
  stride = min(idx[p], jdx[q])
  @finch(@∀ i ∈ step:stride (C[] += val[p] * wal[q]))
  p += stride == idx[p]
  q += stride == idx[q]
  step = stride
end
```

***Jumpers.*** Finch lowers jumpers similarly to steppers. Instead of using the smallest declared extent, jumpers use the largest. Whether the smallest or largest extent is chosen, it will correspond to the exact range of at least one child looplet, which can be lowered verbatim. Choosing the largest extent enables powerful optimizations. If the largest looplet is a run of zeros, multiple child looplets that it multiplies can be skipped. The body of a jumper should be able to process more than one child looplet, but usually includes a switch that specializes to the case where only a single child is needed.

As an example, a jumper over spikes usually declares the length of the current spike but, if another spike is longer, processes multiple spikes with a stepper. The jumper allows us to implement leader-follower or galloping intersections by electing themselves as leaders and truncating other looplets to the largest extent declared by a jumper:

```
A = Jumper(stride = idx[p], body = val[p], next = p += 1)
B = Jumper(stride = jdx[q], body = wal[q], next = q += 1)
p = q = 1
@finch(@∀ i ∈ start:stop (C[] += A[i] * B[i]))
⇓
p = q = 1
step = start
while step < stop
  stride = max(idx[p], jdx[q])
  if stride == idx[p] && stride == jdx[q]
    @finch(@∀ i ∈ step:stride (C[] += val[p] * wal[q]))
    p += 1
    q += 1
  elseif stride == idx[p]
    @finch(@∀ i ∈ step:stride (C[] += val[p] * B[i]))
    p += 1
  elseif stride == idx[p]
    @finch(@∀ i ∈ step:stride (C[] += A[i] * wal[q]))
    q += 1
  end
  step = stride
end
```

***Shifts.*** Shift looplets do not need a special compiler pass, but instead shift the declared extents of their arguments during other passes. Any looplet such as a run, scalar, or spike which results in a terminal scalar value with respect to the loop index can safely discard the shift looplet wrapper.

### 6.2 Choosing Lowerers

The same expression may contain several different looplets, each of which can lowered with a different looplet pass. We use an extensible pairwise pass resolution heuristic. Each outer looplet declares a style, which signals the kind of compiler pass that can lower it. We choose between styles with pairwise resolution rules. When we define a resolution between two styles, we are declaring that the resulting compiler pass can handle all of the looplets that might declare either style. For example, a run style and a spike style can resolve to a spike style, because a spike lowerer can handle runs, while the run lowerer cannot handle spikes. Compiler pass resolution allows us to implement our lowering in an extensible way, where additional lowering passes can be added to the system and incompatibilities are recognized when resolution rules cannot be found.

Our resolution rules also represent heuristics for the order in which looplets should be lowered. Finch chooses between lowering passes in the following order of descending priority:

```
Switch > Run > Spike > Pipeline > Jumper > Stepper > Lookup
```

Our reasoning is as follows: We always lower switch looplets first in order to examine their cases. We lower runs and spikes whenever we see them, to simplify expressions as early as possible. Then, we lower pipelines before the looping constructs to hoist the control flow outside of loops. We lower Jumpers before steppers to give them leader privileges. Finally, if our expression is just lookups, there's nothing left to do but emit a simple for-loop. As future work, we plan to investigate modifying the looplet lowering order by giving some looplets customizable numeric priorities.

## 7 Advanced Protocols

The same level format can be traversed using different protocols. For example, a sorted list format can be randomly accessed with binary search, or the indices could be visited iteratively in ascending order, with different asymptotic effects in different situations [3]. Looplets allow us to express new protocols for existing formats.

For example, Figure 6a unfurls a sparse list with a leader protocol. The outermost jumper declares that this list will have priority during coiteration, and other iterators will follow this list's step size. Merging two lists with a leader protocol enables a galloping (mutual lookahead) intersection, where each list agrees to use the larger step size.

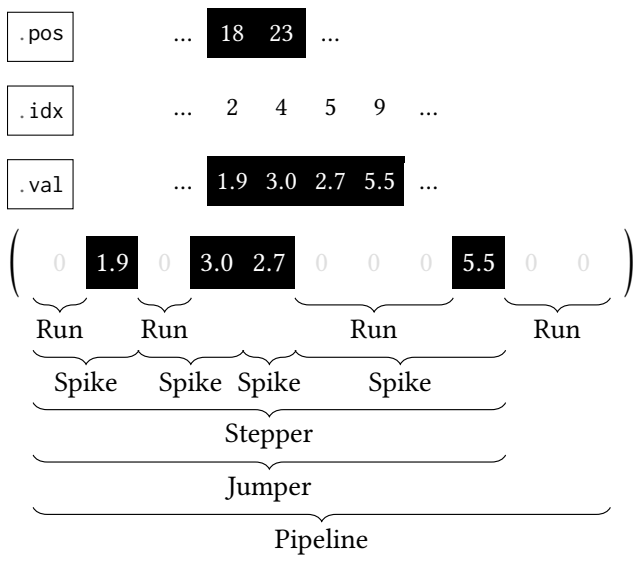

```
Pipeline(
  Phase(
    stride = idx[pos[i+1]-1],
    body = begin
      p = pos[i]
      Jumper(
        seek(j) = (p = search(idx, j)),
        stride = idx[p],
        body(j) = Switch(
          Case(
            cond = idx[p] == j,
            body = Spike(
              body = Run(0)
              tail = val[p])),
          Case(
            body = Stepper(
              seek(j) =
                (p = search(idx, j)),
              stride = idx[p],
              body = Spike(
                body = Run(0)
                tail = val[p]),
              next = p += 1))),
        next = p += 1)
    end),
  Phase(
    body = Run(0)))
```

**(a)** A galloping (leader) protocol for a list format. Compare to the walking (follower) protocol of Figure 3d.

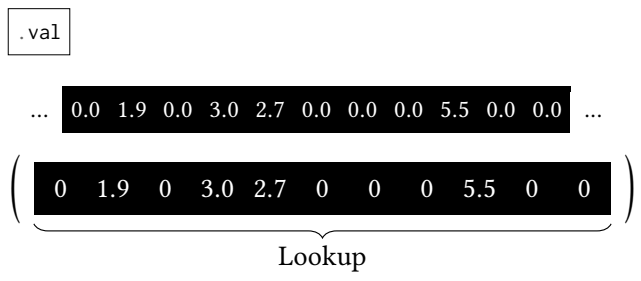

```
Pipeline(
  Lookup(
    body(j) = val[i*n+j]))
```

**(b)** A locate protocol for a dense format. All entries are treated as if they might be nonzero.

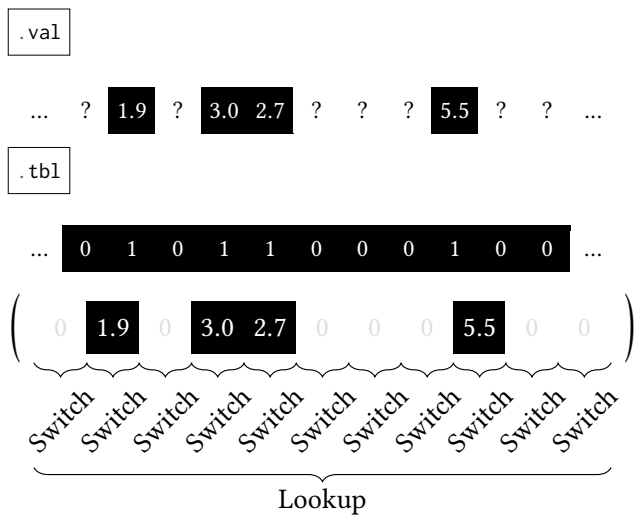

```
Pipeline(
  Lookup(
    body(j) = Switch(
      Case(
        cond = tbl[i*n+j],
        body = val[i*n+j]),
      Case(0))))
```

**(c)** A locate protocol for a bitmap format. The switch branches on whether each value is statically zero.

**Figure 6.** A protocol language also allows us to iterate over the same structure (or even the same format) in different ways, enabling new optimization opportunities.

Figures 6b and 6c compare two strategies for randomly accessing a dense array. However, the latter approach introduces a check for zero, allowing the compiler to specialize for the zero case.

# 8 Index Modifiers

Looplets enable new functionality previously unsupported by sparse array compilers. We show how the combination of few simple **index modifiers** can be combined to implement kernels like concatenation and convolution over sparse inputs. These index modifiers change the behavior of their corresponding mode by wrapping or modifying the looplets that mode would unfurl into.

As an example, consider the special windowing array `window(i, j)[k] = i+k-1` with dimension `1:j-i`. A `window` can be used to represent a slice of an input array. As such, `A[window(3,5)[k]]` would behave like the slice `A[3:5][k]`. We can construct a protocol for `A[window(i,j)[k]]` as

```
Shift(delta=i,body=truncate(unfurl(A), i:j))
```

Letting `offset(i)[j] = j-i` be a special array that shifts the dimension of the parent, we can construct the protocol

```
Shift(delta=i,body=unfurl(A))
```

Finally, we can also introduce a padding array, which allows out-of-bounds access. `permit[i] = i` when `i` is in bounds, but `missing` when out of bounds. Note that `missing` propagates, so `A[missing] = missing` and `f(x, missing) = missing`. We define a function `coalesce` to return it's first non-`missing` argument. Our protocol for `A[permit[i]]` is

```
Pipeline(
  Phase(stride=0, Run(missing)),
  Phase(stride=length(A), body=unfurl(A)),
  Phase(Run(missing)))
```

Together, these primitives greatly expand the range of functionality that the concrete index notation can express. For example, we can concatenate `A` and `B` to produce `C` with:

```
@∀ i C[i] = coalesce(A[permit[i]], B[permit[offset(size(A))[i]]])
```

Similarly, we can express one-dimensional convolution with a vector `A` and a length-3 filter `F` to produce `B` with:

```
@∀ i j B[i] += coalesce(A[permit[offset(2-i)[j]]], F[permit[j]])
```

# 9 Evaluation

We implemented[1] Finch in Julia v1.8.0. All timings are the minimum of at least 10,000 runs or at least 5 seconds of execution time. Kernels were run on an Intel®Xeon®CPU E5-2680 v3 running at 2.50GHz with AVX2 instructions. Turbo Boost was turned off to avoid thermal throttling, and all kernels ran on a single processor.

## 9.1 SpMSpV

Many of the new functionalities introduced by Finch involve coiteration between two structured operands. To emphasize the effects of different coiteration strategies, we begin with a comparison between SpMSpV approaches in Figure 7. Our kernel was `@∀ i j y_ref[i] += A[i, j] * x[j]`. We iterate over $j$ in the inner loop to test coiteration capabilities, repeatedly merging $x$ with every row of $A$. We tested on matrices in the Harwell-Boeing collection [15]. When $x$ had many nonzeros, A leader protocol for $A$ performed well, as it visited each element of $A$ and fast-forwarded $x$. The follower protocol for $A$ led to a few big speedups when the vectors were very sparse. Our matrices are from the scientific computing domain, and frequently contain dense blocks or bands, so we also tried the VBL format, which holds multiple dense bands. Finch processes the index of each band, rather than the index of each element in each band. This advantage became clear when $x$ was very sparse, and VBL led to large speedups over TACO. Galloping proved robust to the relative sparsity differences of the inputs.

[1] https://github.com/willow-ahrens/Finch.jl

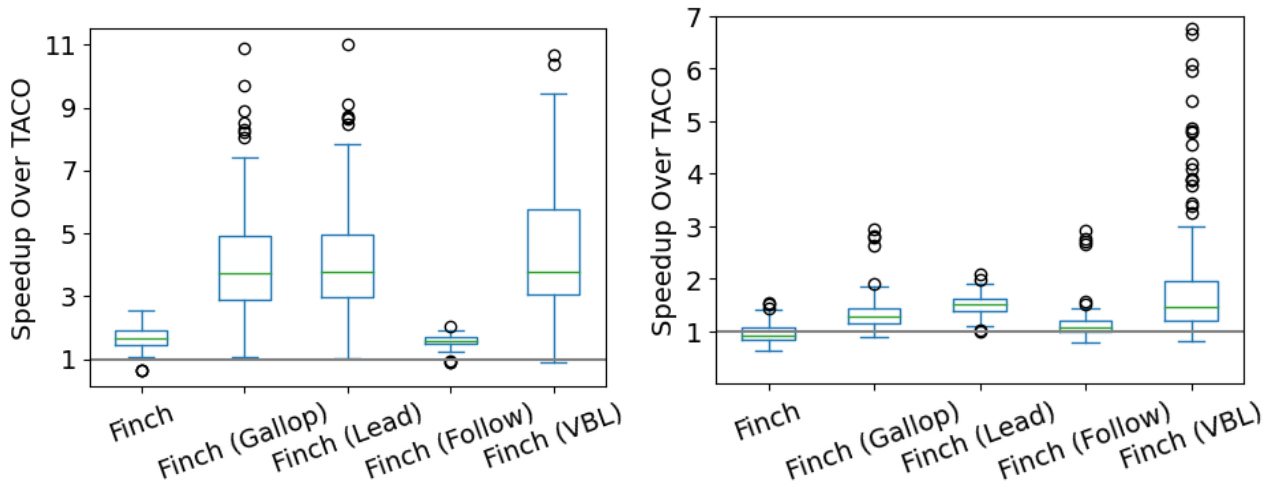

**(a)** $x$ has 10% fraction nonzero. **(b)** $x$ has count of 10 nonzeros.

**Figure 7.** Speedups on SpMSpV with randomly placed nonzeros in $x$. The boxes display quartiles, and the whiskers and outliers display extrema. We tested on all members of the Harwell-Boeing collection with at least 1000 nonzeros [15]. The right plot has one VBL point above the plot area.

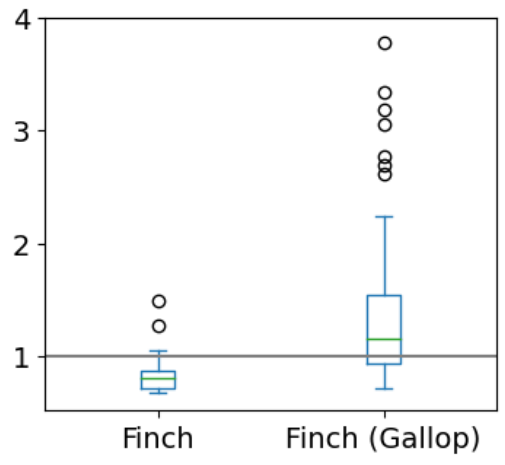

**Figure 8.** Triangle counting Speedups over TACO. The boxes display quartiles, and the whiskers and outliers display extrema. We tested on all members of of the SNAP network dataset collection with less than 1,632,803 vertices [34].

## 9.2 Triangle Counting

Galloping intersections can greatly accelerate the triangle counting kernel, where no loop ordering can avoid an intersection in an inner loop and operands often have unpredictable power-law sparsity distributions of nonzeros within the rows and columns. Our kernel is

```
@∀ i j k C[] += A[i,j] && A[j,k] && A[k,i]
```

Both TACO and Finch transpose the last argument before benchmarking the kernel. Figure8 evaluates our naive two-finger merge and a linear galloping intersection with respect to TACO. Our galloping intersection resulted in order-of-magnitude speedups. Our two-finger merge is not quite as fast as TACO's, indicating that there are still many opportunities for optimization.

## 9.3 Convolution

Protocols enable new sparse functionality. Figure 9 compared our sparse convolution kernel to OpenCV. Our Finch kernel for a masked 11x11 convolution was

```
@∀ i k j l C[i, k] += (A[i, k] != 0) * coalesce(A[permit[offset[6-i, j]],
↪  permit[offset[6-k, l]]], 0) * coalesce(F[permit[j], permit[l]], 0)
```

where we use a binary search to seek to the start of each considered window of A. We consider zero-padded arrays, but our index modifiers can express other padding schemes, such as circular padding by adding copies of $A$ on each side of the original. The binary search allows us to scale the kernel linearly with sparsity of the input. We can think of a sparse convolutional kernel as a form of neighbor counting in a grid of points. Our experiment shows that the sparse implementation begins to outperform the dense one at around 5% sparsity, and results in a 9.5× speedup at 1% sparsity.

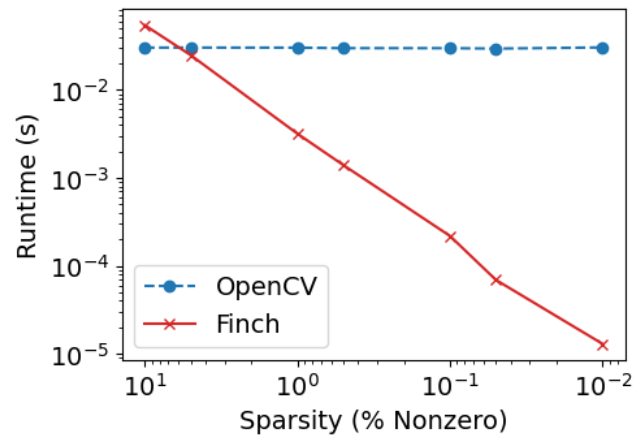

**Figure 9.** Dense versus sparse convolution runtime as the sparsity increases. 1000x1000 randomly sparse floating point grid with dense 11x11 floating point kernel.

## 9.4 Alpha Blending

Finch is competitive with frameworks that move beyond sparsity. Figure 10 compares against both OpenCV and TACO's RLE extensions [14] for an alpha blending kernel

```
@∀ i j A[i, j] = round(UInt8, alpha * B[i, j] + beta * C[i, j])
```

The Omniglot dataset has 105 × 105 images with grayscale handwritten characters from 50 different languages [32]. The Humansketches dataset has 1111 × 1111 images with hand drawn grayscale sketches [16]. Finch's RLE format is competitive on both datasets, though there was not enough structure in Omniglot for an RLE approach to outperform OpenCV.

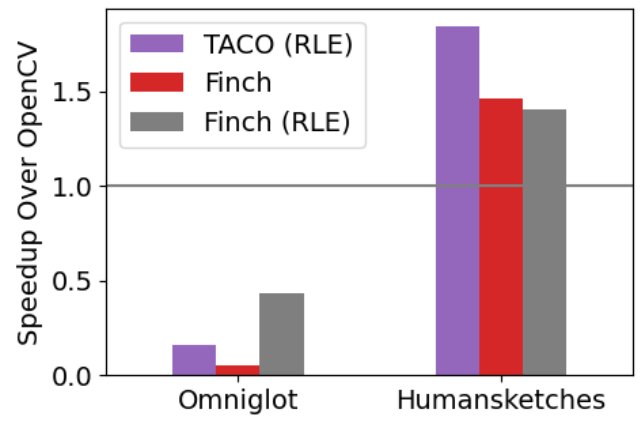

**Figure 10.** We compare TACO's prototype RLE extensions [14] with Finch's sparse and RLE approaches on an alpha blending task. Results are the mean of 10 images.

## 9.5 All-Pairs Image Similarity

Finch enables unique optimizations. Figure 11 considers the effectiveness of different strategies on an all-pairs image similarity kernel. We consider only pairwise comparisons, rather than batch approaches, to focus on coiteration and maintain relevance to k-means clustering methods. The kernel is

```
@∀ k ij R[k] += A[k, ij]^2
@∀ k l ((O[k,l] = sqrt(R[k] + R[l] - 2 * o[])) where (@loop ij o[] += A[k, ij]
↪  * A[l, ij]))
```

Where A contains linearized images in each row. The MNIST dataset contains 28 × 28 images of handwritten digits [33]. EMNIST is an extension which contains similar images [12]. Results compute distances between 256 images. Finch's VBL format can take advantage of the white background and clustered nonzeros in most of these images. However, the Omniglot dataset has noisier backgrounds, which are better captured by run-length-encoding. Additionally, when both accesses to A contain a run, we can apply the last rule of Figure 5 to sum the whole run at once. None of the approaches were able to beat OpenCV on this processor, which uses vector units on the 8-bit image data. Future work might investigate whether quantization induces more structure.

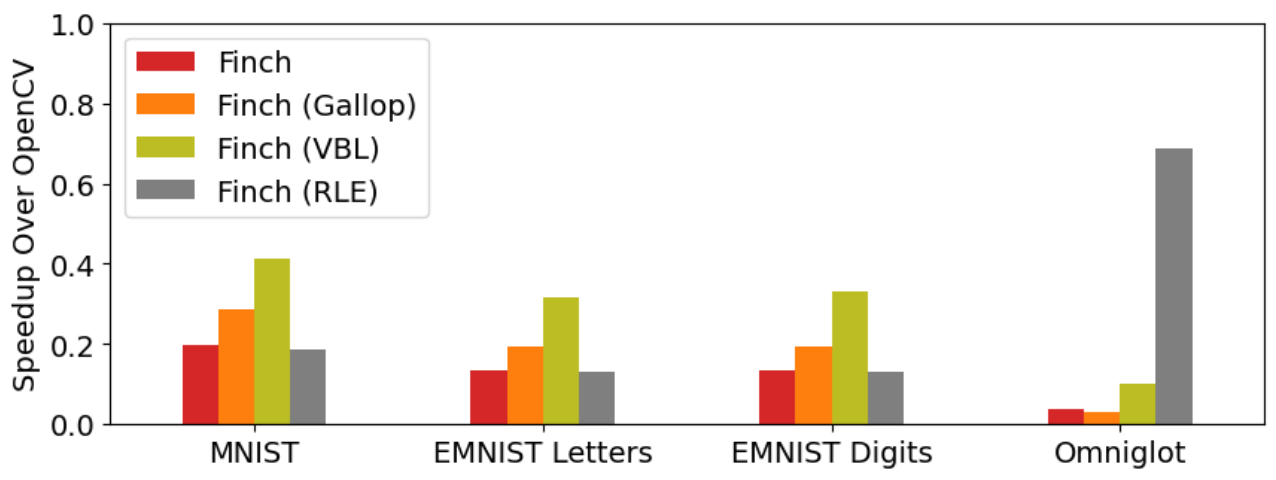

**Figure 11.** Speedups on all pairs image similarity.

## 10 Related Work

Dense array frameworks like APL [22], BLAS [10], and NumPy [18] greatly influenced subsequent language design.

The Halide compiler popularized array optimization through the use of scheduling commands, or semantics-preserving program transformations [35, 42, 43]. The scheduling transformations within TACO are performed on the same language (CIN) as ours, and implementing a scheduling language in Finch is future work.

Array compilers that support irregular sparsity include TACO [25, 26], MT1 [6, 8, 9], MLIR [7], COMET[51], Etch [31], SIPR [41], Tiramisu [4] and CHiLL-I/E [48].

Some approaches specialize for particular sparsity patterns. The OSKI library [53] includes specializations for block sparsity, and the BLAS for triangular matrices. TACO supports fixed-size blocks and bands[11]. TESA specializes for bitmap or quantized sparsity[57], CORA for ragged arrays [17], and TAICHI for spatial sparsity [21]. SparseTIR supports combinations of multiple formats [55]. Zhao et. al. extend the sparse polyhedral framework to support coiteration for conjuctive (*) leader-follower loops over a wide variety of formats, but do not support disjunction (+) or mutual lookahead (galloping) coiterations [56].

Most sparse frameworks support only numeric types and the + or · operators, but supporting arbitrary element types and operators can transform such frameworks into productive programming models. TACO [19], GraphBLAS [13, 23, 37], and Cyclops [46] have been extended to support new element types and operators.

Previously mentioned sparse compilers consider sparse arrays as a set of nonzeros to be processed, precluding new optimizations. Compilers like StreamIt[50] and an extension to TACO[14] support direct computation on losslessly compressed datasets. The BLAS and Cyclops framework both optimize for dense symmetry [10, 47].

Many approaches model sparse computation with database queries, including the Bernoulli compiler [28–30] and a TACO autoscheduler [3]. Run-length encoding is a popular format for column stores [38]. Queries can be modeled and optimized as iterators [39].

In functional programming, stream fusion is a related technique which can fuse lazily constructed streams [24, 36].

Several sparse compilers have been extended to better adapt computation to multicore, GPU, or distributed architectures [4, 44, 46, 54]. Hsu et. al. investigate sparse compilation for spatial-dataflow accelerators [20]. Future work includes the use of our technique to more easily target new architectures.

## 11 Conclusion

Historically, specializations for the array structures in different fields have been handled separately. For example, scientific computing specializes for block sparsity, and image processing specialized for compression techniques. Our work takes a step towards unifying these techniques across disciplines, providing opportunities to transfer optimizations and better specialize to data with heterogeneous structures.

## Acknowledgments

This work was supported by NSF Grant IIP-2044424 and the Applications Driving Architectures (ADA) Center, a JUMP Center cosponsored by SRC and DARPA.